%% file: main.tex
\documentclass[a4paper,UKenglish]{lipics-v2016}

\usepackage{microtype}
\usepackage{setspace,
            picins
            }
\usepackage{extarrows}
\usepackage{color}
\usepackage{extarrows}
\usepackage{multirow}
\usepackage[table]{xcolor}

\newcommand{\lcolorbox}[2]{%
\hspace*{-\fboxsep}\colorbox{#1}{#2}%
}

\title{Proving Linearizability via Branching Bisimulation\footnote{This work was supported by NSFC 61100063 and Alexander von Humboldt.}}

\author[1,2]{Xiaoxiao Yang}
\author[2]{Joost-Pieter Katoen}
\author[1]{Huimin Lin}
\author[2]{Hao Wu}
\affil[1]{State Key Laboratory of Computer Science, Institute of Software, Chinese Academy of Sciences, Beijing, China}
\affil[2]{Software Modeling and Verification, RWTH Aachen University, Germany}
\authorrunning{X. Yang, J. -P. Katoen, H. Lin and H. Wu} 
\Copyright{Xiaoxiao Yang, Joost-Pieter Katoen, Huimin Lin, Hao Wu}
\subjclass{please refer to \url{http://www.acm.org/about/class/ccs98-html}}

\ArticleNo{13}

\begin{document}

\input{macro}

\maketitle

\begin{abstract}
Linearizability and progress properties are key correctness notions for concurrent
objects. However, model checking linearizability has suffered from the PSPACE-hardness
of the trace inclusion problem.
This paper proposes to exploit branching bisimulation, a fundamental semantic
equivalence relation developed for process  algebras which can be computed efficiently,
in checking these properties. A quotient construction
is provided which results in
huge state space reductions. We confirm the advantages of the proposed approach
on more than a dozen benchmark problems.
 \end{abstract}

\input{intro}

\input{objsys}

\input{co-bis}
\input{divergence}

\input{experiments}

\input{relatedwork}

\vspace{3mm}

\noindent
{\bf Acknowledgement} We thank the CADP support team for their helps and
patience during the experiments.


\pagebreak

\input{appendix}


\end{document}

%% file: macro.tex
\newtheorem{Def}{Definition}
\newtheorem{Expl}{Example}
\newtheorem{Thm}{Theorem}
\newtheorem{Lem}[Thm]{Lemma}
\newtheorem{fac}[Thm]{Fact}
\newtheorem{Cor}[Thm]{Corollary}

\def\squareforqed{\hbox{\rlap{$\sqcap$}$\sqcup$}}
\def\qed{\ifmmode\squareforqed\else{\unskip\nobreak\hfil
\penalty50\hskip1em\null\nobreak\hfil\squareforqed
\parfillskip=0pt\finalhyphendemerits=0\endgraf}\fi}

\newcounter{statement}
\def\stmnum{\hbox to .01pt{}\rlap{\rm \hskip -\displaywidth\thestatement.}}
\def\stm{\refstepcounter{statement}\topsep 2pt \trivlist \item[]\leavevmode
\hbox to\linewidth\bgroup $ \displaystyle \hskip\leftmargini}
\def\endstm{$\hfil \displaywidth\linewidth\stmnum\egroup \endtrivlist}

\newenvironment{Proof}{%
\begin{list}{}{\setlength{\topsep}{\jot}\setlength{\parsep}{\topsep}%
\addtolength{\parsep}{-0.3\parsep}\setlength{\leftmargin}{0pt}}%
\parindent 4ex
\item[]\setcounter{statement}{0}\textbf{Proof:}}{\end{list}}

\newcommand{\scarrow}[2]
      { \mathrel{\setbox0 \hbox{ ${\scriptstyle #1}$ }\displaystyle
            \mathop{\hbox to \wd0{$\Longrightarrow$}}\limits_{#1}^{#2}}\,\,\,
      }
\newcommand{\lscarrow}[2]
      { \mathrel{\setbox0 \hbox{ ${\scriptstyle #1}$ }\displaystyle
                 \mathop{\hbox to \wd0{$=\!=\!\Longrightarrow$}}\limits_{#1}^{#2}}
      }
\newcounter{LLN}
\newcommand{\beginit} {
                         \begin{list}{\hfill{\rm \arabic{LLN}.} }{\usecounter{LLN}
                         \setlength{\leftmargin}{\leftmarginii}
                         \setlength{\itemsep}{\smallskipamount}}}

\newbox\mystrutbox
\setbox\mystrutbox=\hbox{\vrule height10pt depth4pt width0pt}
\def\mystrut{\relax\ifmmode\copy\mystrutbox\else\unhcopy\mystrutbox\fi}
\newcommand{\lprf}[3]
       {\llap{{$#3$}\thinspace}{{\mystrut\displaystyle #1}
          \over
         {\mystrut\displaystyle #2}}}
\newcommand{\rprf}[3]
       {{{\mystrut\displaystyle #1}
          \over
         {\mystrut\displaystyle #2}}{\rlap{{$#3$}\thinspace}}}

\newcommand{\dotminus}{-\!\!\!\!^{\textstyle.}\!\!\!\!-}
\newcommand{\lit}[1]{{\rm [#1]}}
\newcommand{\henv}[1]{#1}
\newcommand{\setof}[2]{\{ #1 \: | \: #2 \}}
\newcommand{\bigsetof}[2]{\bigl\{ #1 \: \bigm| \: #2 \bigr\}}
\newcommand{\arrow}[1]{\stackrel{#1}{\longrightarrow}}
\newcommand{\sarrow}[1]{\stackrel{#1}{\Longrightarrow}}
\newcommand{\dobarrow}[1]{\stackrel{#1}{\Longrightarrow}}
\newcommand{\proc}{Pr}
\newcommand{\af}[1]{\forall{\rm F\;}#1}
\newcommand{\ef}[1]{\exists{\rm F\;}#1}
\newcommand{\ag}[1]{\forall{\rm G\;}#1}
\newcommand{\eg}[1]{\exists{\rm G\;}#1}
\newcommand{\varu}[3]{U^{#1}_{#2,#3}}
\newcommand{\varusfg}{\varu{s}{F}{G}}
\newcommand{\varuwfg}{\varu{w}{F}{G}}
\newcommand{\semu}[3]{{\cal U}^{#1}_{#2,#3}}
\newcommand{\semuwfg}{\semu{w}{F}{G}}
\newcommand{\semusfg}{\semu{s}{F}{G}}
\newcommand{\transu}[3]{{\cal T}^{#1}_{#2,#3}}
\newcommand{\transusfg}{\transu{s}{F}{G}}
\newcommand{\transuwfg}{\transu{w}{F}{G}}
\newcommand{\comp}[1]{{\cal C}(#1)}
\newcommand{\prc}[1]{{\cal P}(#1)}
\newcommand{\comppred}[1]{\prec_{#1}}
\newcommand{\bisim}[1]{\sim_{#1}}
\newcommand{\fat}[1]{\mbox{\bf #1}}
\newcommand{\tr}{\mbox{\rm tt}}
\newcommand{\fa}{\mbox{\rm ff}}
\newcommand{\act}{Act}
\newcommand{\impliess}{\: \Rightarrow \:}
\newcommand{\implied}{\Leftarrow}
\newcommand{\implieds}{\:\Leftarrow\:}
\newcommand{\iffs}{\:\Leftrightarrow\:}
\newcommand{\hviss}{\Leftrightarrow}
\newcommand{\sigent}{\sigma_{\entail{}}}
\newcommand{\ua}[2]{#1 \subseteq \sem{\D}#1 \cup #2}
\newcommand{\id}{\mbox{\rm Id}}
\newcommand{\powerset}[1]{{\Large \wp} (#1)}
\newcommand{\sem}[1]{\lbrack\!\lbrack #1 \rbrack\!\rbrack}
\newcommand{\abs}[1]{|\!| #1 |\!|}
\newcommand{\may}[1]{\langle #1 \rangle}
\newcommand{\wmay}[1]{\langle\!\langle #1 \rangle\!\rangle}
\newcommand{\wmust}[1]{\sem{#1}}
\newcommand{\until}[1]{[ #1 \rangle}
\newcommand{\smay}[1]{\langle\!\cdot #1 \cdot\!\rangle}
\newcommand{\must}[1]{[ #1 ]}
\newcommand{\smust}[1]{\lbrack\!\cdot #1 \cdot\!\rbrack}
\newcommand{\sat}[1]{\models_{#1}}
\newcommand{\mx}{{\rm max}}
\newcommand{\mn}{{\rm min}}
\newcommand{\satmn}{\sat{\mn}}
\newcommand{\satmx}{\sat{\mx}}
\newcommand{\sigmax}{\sigma_{\mx}}
\newcommand{\sigmin}{\sigma_{\mn}}
\newcommand{\entail}[1]{\vdash_{#1}}
\newcommand{\entmn}{\entail{\mn}}
\newcommand{\entmx}{\vdash}
\newcommand{\satt}[1]{|\!\!\!\equiv_{#1}}
\newcommand{\sattmx}{\satt{\mx}}
\newcommand{\sattmn}{\satt{\mn}}
\newcommand{\Dtu}{\D^{\tau}}
\newcommand{\Dtd}{\D_{\tau}}
\newcommand{\da}[2]{\semd #1 \cap #2 \subseteq #1}
\newcommand{\ass}[2]{#1 : #2}
\newcommand{\MM}[1]{{\cal M}_{#1}}
\newcommand{\D}{{\cal D}}
\newcommand{\mmid}{{\cal M}_{{\rm Id}}}
\newcommand{\og}{\wedge}
\newcommand{\eller}{\vee}
\newcommand{\semd}{\sem{\D}}
\newcommand{\M}{{\cal M}}
\newcommand{\infrule}[3]
           {\parbox{2cm}{ $$ {\frac {#1}{#2}}\hspace{.5cm}{#3} \hfill $$}}
\newcommand{\infrulegen}[4]
           {{#1}\hspace{.5cm}{\frac {#2}{#3}}\hspace{.5cm}{#4}}
\newcommand{\ent}{\entail{}}
\newcommand{\Gammah}{\widehat{\Gamma}}
\newcommand{\inv}[1]{\textsf{\rm Inv}(#1)}
\newcommand{\pos}[1]{\mbox{\rm Pos}(#1)}
\newcommand{\inva}{\inv{\may{a}\tr}}
\newcommand{\posa}{\pos{\must{a}\fa}}
\newcommand{\op}{{\cal O}}
\newcommand{\even}[1]{\mbox{\rm Even}(#1)}
\newcommand{\unic}{\bigr( \lbr\;|\;\hole\bsl p\bigr) \bsl\coin,\cof}
\newcommand{\live}{\mbox{\rm Live}}
\newcommand{\con}{\mbox{\rm Con}}
\newcommand{\com}{\mbox{\rm Com}}
\newcommand{\scom}{\mbox{{\cal Sc}}}
\newcommand{\dead}{\mbox{\rm Dead}}
\newcommand{\diver}{\mbox{\rm Div}}
\newcommand{\sUntil}[2]{\mbox{\rm Unt}^s(#1,#2)}
\newcommand{\wUntil}[2]{\mbox{\rm Unt}^w(#1,#2)}
\newcommand{\saf}[1]{\mbox{\rm Saf}(#1)}
\newcommand{\uni}{\mbox{\sl Uni}}
\newcommand{\staff}{\mbox{\sl Staff}}
\newcommand{\equip}{\mbox{\sl Equip}}
\newcommand{\acc}{\mbox{acc}}
\newcommand{\del}{\mbox{del}}
\newcommand{\wip}[2]{\mbox{\rm wip}(#1,#2)}
\newcommand{\sop}[2]{\mbox{\rm sop}(#1,#2)}
\newcommand{\cuni}{C_{\mbox{\rm uni}}}
\newcommand{\lbr}{\mbox{\rm lbr}}
\newcommand{\carrow}[2]{\raisebox{0.2ex}{$\,\,\,\,\,{#2\atop #1}
\!\!\!\!\!\!\!\!\arrow{}\,\,
$}}
\newcommand{\ccarrow}[2]{\raisebox{0.2ex}{$\,\,\,\,\,\,{#2\atop #1}
\!\!\!\!\!\!\!\!\!\arrow{}\!\!\!\!\!\!\!\!\arrow{}\,\,
$}}
\newcommand{\dccarrow}[2]{\raisebox{0.2ex}{$\,\,\,\,\,\,{#2\atop #1}
\!\!\!\!\!\!\!\!\!\arrow{}\!\!\!\!\!\!\!\!\arrow{}_\C\,\,
$}}
\newcommand{\cdarrow}[2]{\raisebox{0.2ex}{$\,\,\,\,\,\,{#2\atop #1}
\!\!\!\!\!\!\!\!\!\arrow{}_\C\,
$}}
\newcommand{\dcarrow}[2]{{\raisebox{-1.2ex}{
                         $\stackrel{#2}{\stackrel{\longrightarrow_\Diamond}
                          {\scriptstyle #1}}$  }}}
\newcommand{\rearrow}[2]{\raisebox{-1.2ex}{
                         $\stackrel{#1}{\stackrel{\longrightarrow}
                          {\scriptstyle #2}}$  }}
\newcommand{\bsl}{\backslash}
\newcommand{\hole}{\mbox{$[\;]$}}
\newcommand{\coin}{\mbox{coin}}
\newcommand{\cof}{\mbox{cof}}
\newcommand{\unicp}{\bigr( \cof .(\lbr + p.\lbr) \;|\;\hole\bsl p\bigr)
                                   \bsl \coin,\cof}
\newcommand{\unicpp}{\bigr( (\lbr + p.\lbr)\;|\;\hole\bsl p\bigr)
                             \bsl \coin,\cof}
\newcommand{\synen}[2]{#1\,\,{\bf{\sf with}}\,\,#2}
\newcommand{\emax}[2]{\mbox{{\bf {\sf max }}}#1\,\,{\bf{\sf with}}\,\,#2}
\newcommand{\emin}[2]{\mbox{{\bf {\sf min }}}#1\,\,{\bf{\sf with}}\,\,#2}
\newcommand{\lmax}{\mbox{{\sf max}\,}}
\newcommand{\lmin}{\mbox{{\sf min}\,}}
\newcommand{\Dsem}[1]{{\sf D} \lbrack\!\lbrack #1 \rbrack\!\rbrack}
\newcommand{\Delsem}[1]{{\sf \Delta} \lbrack\!\lbrack #1 \rbrack\!\rbrack}
\newcommand{\Asem}[1]{{\sf E} \lbrack\!\lbrack #1 \rbrack\!\rbrack}
\newcommand{\Lsem}[1]{{\sf L} \lbrack\!\lbrack #1 \rbrack\!\rbrack}
\newcommand{\Nsem}[1]{{\sf N} \lbrack\!\lbrack #1 \rbrack\!\rbrack}
\newcommand{\DMTSt}{\mbox{DMTS$^2$}\,}
\newcommand{\DMTSs}{\mbox{DMTS$^\ast$}\,}
\newcommand{\conf}[2]{\langle #1,#2\rangle}
\newcommand{\chop}[1]{\mbox{{\tt chop}$(#1)$}}
\newcommand{\tl}[1]{\mbox{{\tt tl}$(#1)$}}
\newcommand{\eemax}[1]{\mbox{{\bf {\sf max }}}#1}
\newcommand{\eemin}[1]{\mbox{{\bf {\sf min }}}#1}
\newcommand{\pre}{\mbox{\,{\tt pre}\,}}
\newcommand{\conft}[1]{\langle #1\rangle}
\newcommand{\incon}{IC}
\newcommand{\hml}{{\cal M}}
\newcommand{\depth}{depth}
\newcommand{\dsat}{{\,\mbox{{\tt sat}}}}
\newcommand{\branb}{{\approx_b}}
\newcommand{\V}{{\cal V}}
\newcommand{\Z}{{\cal Z}}
\newcommand{\ins}[3]{\vdash^?_{#3}#1\colon#2}
\newcommand{\insp}[3]{\vdash^{#3}#1\colon#2}
\newcommand{\insn}[3]{\not\vdash_{#3}#1\colon#2}
\newcommand{\wi}[2]{{\cal W}(#1,#2)}


\newcommand{\DEF}               {\stackrel{\rm def}{=}}
\newcommand{\invo}              {\textsf{inv}}
\newcommand{\res}               {\mathit{res}}
\newcommand{\lin}               {\textsf{lin}}
\newcommand{\Abs}               {\textsf{Abs}}
\newcommand{\client}            {\textsf{Client}}
\newcommand{\Sys}               {\textsf{Sys}}
\newcommand{\Systau}            {\textsf{Sys1}}
\newcommand{\Sysinv}            {\textsf{Sys2}}
\newcommand{\Sysres}            {\textsf{Sys3}}
\newcommand{\Sysobj}            {\textsf{Sys4}}
\newcommand{\emp}               {\varepsilon}
\newcommand{\Ex}                {\bf Ex}
\newcommand{\Eva}               {\bf Eva}
\newcommand{\true}              {\mathit{true}}
\newcommand{\false}             {\mathit{false}}

\newcommand{\StackOp}           {\mathit{StackOp}}
\newcommand{\LesOp}             {\mathit{LesOp}}
\newcommand{\TryStackOp}        {\mathit{TryStackOp}}
\newcommand{\TryCollision}      {\mathit{TryCollision}}
\newcommand\ignore[1]{}
\newcommand{\call}              {\texttt{call}}
\newcommand{\ret}               {\texttt{ret}}

\newcommand{\xhookrightarrow}[1]  {\stackrel{#1}{\hookrightarrow}}


\definecolor{ablue}{RGB}{0,153,225}
\definecolor{aorange}{RGB}{225,64,0}
\definecolor{apurple}{RGB}{163,143,196}
\definecolor{corg}{RGB}{204,102,0}

\newcommand{\xx}[1]{\textcolor{ablue}{\bf{#1}}}
\newcommand{\jp}[1]{\textcolor{aorange}{\bf{#1}}}

\newcommand{\atl}[1]{\texttt{\small{Line #1}}}
\newcommand{\atls}[1]{\texttt{\small{Lines #1}}}

\newenvironment{narrow}[2]{%
\begin{list}{}{%
\setlength{\topsep}{0pt}%
\setlength{\leftmargin}{#1}%
\setlength{\rightmargin}{#2}%
\setlength{\listparindent}{\parindent}%
\setlength{\itemindent}{\parindent}%
\setlength{\parsep}{\parskip}}%
\item[]}{\end{list}}

%% file: intro.tex
\section{Introduction}

\newcommand{\forget}[1]{}
\newcommand{\dar}{\Longrightarrow}

A concurrent data structure, or a concurrent object,
provides a set of methods 
that allow client threads to simultaneously access and
manipulate a shared object.
\emph{Linearizablity}~\cite{Herlihy90}
is a widely accepted correctness criterion for
implementations of concurrent objects.
Intuitively, an implementation of a concurrent object is \textit{linearizable} with respect to a sequential specification
if every method call appears ``to take effect'', {\it i.e.} changes the
state of the object, instantaneously
at some time point between its invocation and its response,
behaving as defined by the specification.
Such a time point, which corresponds to the execution of some program statement,
is referred to as the {\em linearization point} of the method call.
The difficulties (and confusions) encountered in verifying linearizability
for concurrent data structures  stemmed from the fact that
the linearization points of different calls of the
same method may correspond to different statements in the method's, or
even other method's, program text.



The subtlety of linearization points can be illustrated using
the heavily studied Herlihy and Wing queue algorithm \cite{Herlihy90},
shown in Figure \ref{pic-6}.
It has two methods, Enq (enqueue) and
Deq (dequeue).
The queue is implemented by an array $AR$ of unbounded length, with $back$ as the index of
the next unused slot in $AR$. Each element of $AR$ is initialized to a special value $null$, and
$back$ is initialized to 1.
An Enq execution contains two steps, 
it first gets a local copy $i$ of $back$ and increments $back$,
then stores the new value at $AR[i]$.
A Deq execution may take several steps to find a non-null element to be
dequeued, by visiting $AR$ in ascending order,
starting from index $1$ and ending at $back-1$. At each slot $i$, the current element $AR[i]$ is swapped with $null$. If Deq finds a non-$null$ value, it
will return that value, otherwise it tries the next slot. If no element is found in the entire array, Deq restarts
the search.  
The Enq and Deq methods can be executed concurrently by any number of client threads.
Every execution step is atomic.

\begin{figure}[htpb]
\begin{minipage}{.70\textwidth}
\vspace{-2ex}
\begin{lstlisting}[basicstyle=\scriptsize\ttfamily]
E0 Enq(x:T) {
E1 (i, back):=(back, back+1);  /* increment */
E2 AR[i]:=x; /* store */
E3 return
E4 }
\end{lstlisting}
\begin{lstlisting}[basicstyle=\scriptsize\ttfamily]
D0 Deq() {
D1  while true do {
D2    range := back;
D3    for (0 < i < range) do {
D4      (x, AR[i]):=(AR[i], null);  /* swap */
D5      if (x != null) then return (x)
D6 } } }
\end{lstlisting}
\end{minipage}
\vspace{-2ex}
\caption{Herlihy and Wing queue.} \label{pic-6} 
\end{figure}

\begin{figure}[h!]
  \centering
  \vspace{-1.8ex}
  \includegraphics[scale=.50]{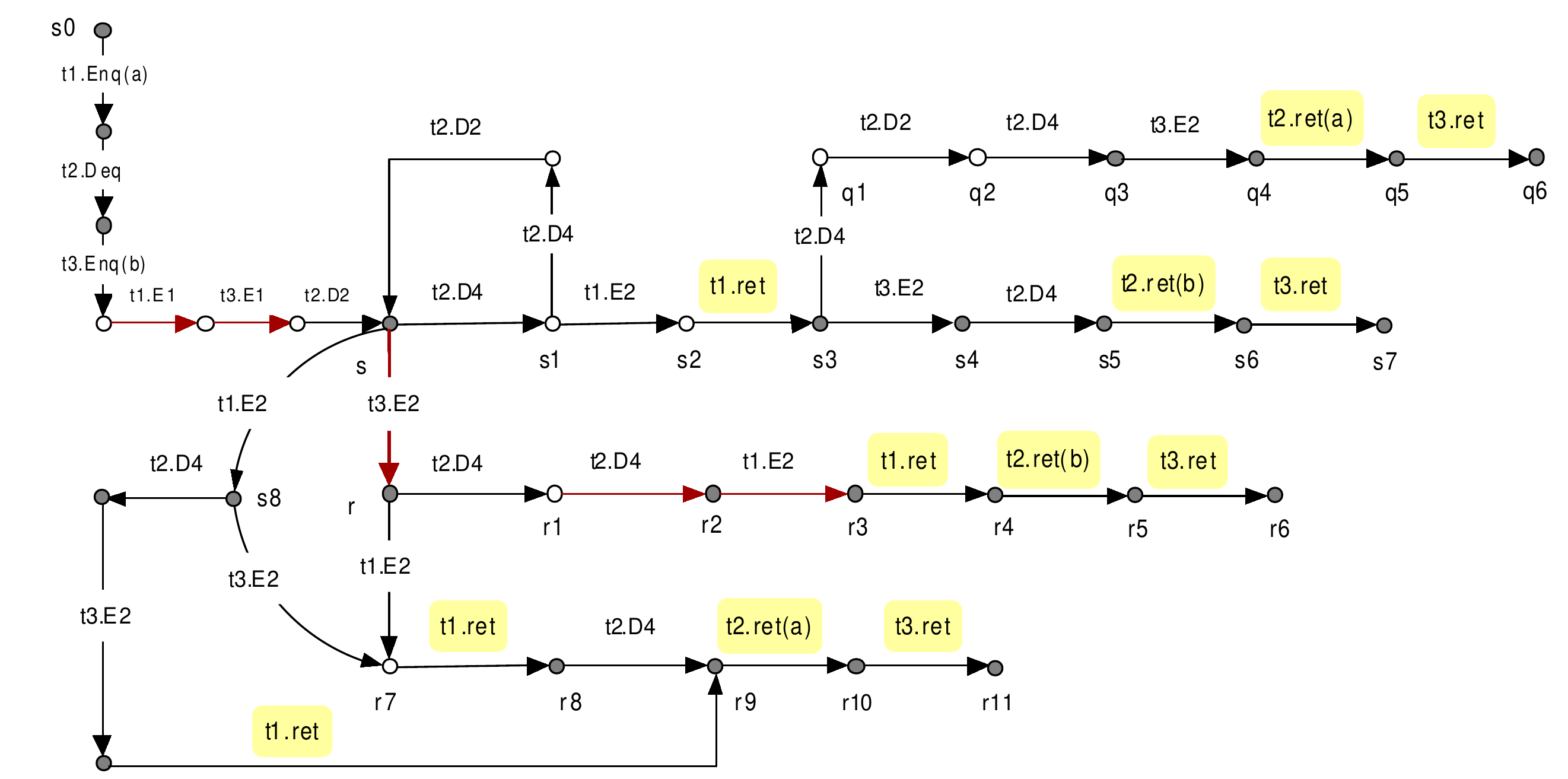}\\
  \caption{A part of the transition system for the Herlihy and Wing queue.}\label{HW-traces}
\end{figure}

The behavior of a concurrent object system
can be modeled as a labeled transition system.
For the HW-queue example,
consider a system of three client threads $t_1$, $t_2$ and $t_3$,
with $t_1$ executing $Enq(a)$, $t_2$ executing $Deq$ and $t_3$ executing $Enq(b)$ concurrently.
A part of the transition graph generated from the system is depicted in Figure \ref{HW-traces},
where $s_0$ is the initial state, and the invocation events
of the $Enq$ and $Deq$ methods (i.e., statements $E0$ and $D0$) 
of a thread $t$ are denoted by $t.Enq(v)$ and $t.Deq()$, respectively.
All internal computation steps of
a method call are regarded as invisible, and labeled with $\tau$.
For the sake of readability, each $\tau$ transition is also marked with the corresponding line number ($E_i$ or $D_i$)
in the program text. A sequence of $\tau$ transitions will be denoted by $\dar$.
The states marked with $\circ$ have some additional transitions which
are irrelevant to the discussions below and hence omitted.

Some linearization points are colored red in the figure.
For instance, $\tau(t_1.E_1)$ is the linearization point for the call of $Enq(a)$ by $t_1$ (starting at $s_0$ and ending at $r_8$)
on the execution trace from $s_0$ to $r_{11}$, since dequeuer $t_2$ first
reads $AR[1]$ then returns $t_2.ret(a)$. However, it is not a linearization point on the trace from $s_0$ to $r_6$, since
the dequeuer $t_2$ first meets the non-$null$ slot at $AR[2]$ and returns $t_2.ret(b)$. Instead, the linearization point of the call of the same method
by $t_1$ on the latter trace is $r_2 \xlongrightarrow{\tau(t_1.E_2)} r_3$.

\forget{
The sequence of transitions
$s_0 \xlongrightarrow{t_1.Enq(a)} \xlongrightarrow{t_2.Deq} \xlongrightarrow{t_3.Enq(b)}$
$\xlongrightarrow{\tau(t_1.E_1)} \xlongrightarrow{\tau(t_3.E_1)} \xlongrightarrow{\tau(t_2.D_2)} s$
in the figure corresponds to the scenario where thread $t_1$  got
slot $AR[1]$, $t_3$ got  slot $AR[2]$, and $t_2$ set its $range$ to $3$.
From $s$ there are three branches, reflecting the possible interleaving orders of $t_1.E_2$,
$t_3.E_2$ and $t_2.D_4$ which
determine whether the value read by thread $t_2$ is from $AR[1]$ or $AR[2]$.

\begin{itemize}
  \item For step $s \xlongrightarrow{t_3.E_2} r$, the traces of $r$ are included in the trace set of $s$, where $t_2$ returns either $t_2.ret(b)$ or $t_2.ret(a)$ in the subsequent
  execution of $r$.

  \item If the transition  $s \xlongrightarrow{t_1.E_2} s_8$ is taken then
        $t_1$ stores $a$ at $AR[1]$
        {\em before} $t_2$ reads  from $AR[1]$.
        Thus $t_2$ always returns $t_2.ret(a)$ regardless of the other interleavings.

  \item If the transition  $s \xlongrightarrow{t_2.D_4} s_1$ is taken then,
       since $t_2$ reads $null$ from $AR[1]$, it will continue to read the next slot $AR[2]$. Thus there are two cases at $s_1$:
        \begin{itemize}
        \item For the path from $s_1$ to $s_7$, the dequeuer $t_2$ reads the non-element from slot $AR[2]$ and returns $t_2.ret(b)$
        \item For the path from $s_1$ to $q_6$, $t_2$ reads $null$ from $AR[2]$ and restart the search from the first slot $AR[1]$ such that
         the dequeuer $t_2$ returns $t_2.ret(a)$.
        \end{itemize}
\end{itemize}
}

An interesting linearization point is
$s \xlongrightarrow{\tau(t_3.E_2)} r$ of the call of $Enq(b)$ by thread $t_3$.
It stores $b$  at $AR[2]$
successfully, changing the empty queue to the queue with just one element $b$,
so that the dequeuer $t_2$ eventually returns $b$
(as witnessed by the $r_4 \xlongrightarrow{t_2.ret(b)} r_5$ transition)
on the trace from $s_0$ to $r_6$.
It is not difficult to see that $s$ and $r$ have
the same set of traces. First, since $s \xlongrightarrow{\tau} r$ and $\tau$ transitions are abstracted away,
every trace of $r$ is also a trace of $s$.
The other direction of inclusion can be seen by observing, for instance, that
the two traces from $s$ below
\begin{quote}
$s \dar s_2 \xlongrightarrow{t_1.ret} s_3 \dar q_4 \xlongrightarrow{t_2.ret(a)} q_5  \xlongrightarrow{t_3.ret} q_6$ and \\
$s \dar s_2 \xlongrightarrow{t_1.ret} s_3 \dar s_5 \xlongrightarrow{t_2.ret(b)} s_6  \xlongrightarrow{t_3.ret} s_7$
\end{quote}
can be matched, respectively, by
the following  traces from $r$
\begin{quote}
$r \dar r_7 \xlongrightarrow{t_1.ret} r_8 \xlongrightarrow{} r_9 \xlongrightarrow{t_2.ret(a)} r_{10}  \xlongrightarrow{t_3.ret} r_{11}$ and \\
$r \dar r_3 \xlongrightarrow{t_1.ret} r_4 \xlongrightarrow{t_2.ret(b)} r_5  \xlongrightarrow{t_3.ret} r_6$
\end{quote}
This is a well-known phenomenon in concurrency: although $s$ and $r$ have the same set of traces,
their behaviors are different because the execution from $s$ branches at $s_3$,
after performing $t_1.ret$, while the execution from $r$
branches at $r$, before performing $t_1.ret$.
Thus \emph{branching potentials play a vital
role in determining linearization points}.

Linearizability can be verified by trace inclusion \cite{Liu13}, which is infeasible
in practice because checking trace inclusion is PSPACE-hard.
The purpose of this paper is to propose a state space reduction technique
based on quotient construction to alleviate the problem. To this end we need to
find a suitable equivalence relation  satisfying the following conditions:
(1) it should have an efficient algorithm, (2) the resulted quotient systems should be
substantially smaller than the original ones,
and (3) it should preserve linearization points.
Conditions (1) and (2) are obvious. Condition (3) is also important because  verification
will be carried out on the quotient systems, thus the diagnoses generated
by verification tools will not be of much help if the information on linearization points
got lost in the quotient construction.

As mentioned before, a linearization point is an internal computation step of
a method call that   ``takes effect'' to change the object's
state.  A common understanding is that an object owns a shared data structure,
and changing its state means changing the value stored in the  data structure.
In the HW-queue algorithm, a queue is represented by two pieces of data: an array $AR$ and an index $back$. An $Enq$ method call modifies
them in two separate steps $E_1$ and $E_2$, which can be interleaved with
the executions of either $Enq$ or $Deq$ methods by other threads.
Which of the two steps actually ``takes effect'' to change the queue's state
can only be determined by the values
later returned by the calls of $Deq$,
as manifested by the visible actions
$ret(a)$ or $ret(b)$ in the example discussed above.
This leads us to take an observational approach.
We need to distinguish between two kinds of $\tau$-steps: those change the overall
state of the transition system, and those do not.
Linearization points belong to the former.
Such distinction is captured by a well-established notion
of behavioral equivalence in concurrency theory --
{\em branching bisimulation},
which preserves computation together with the branching potentials of
all intermediate states that are passed through.
As a consequence, two branching bisimilar states have the same observational
behavior along not only  ordinary traces but also traces at any higher levels~\cite{Glabbeek96}.
Moreover, branching bisimulation can be computed
efficiently \cite{Groote90,DBLP:conf/tacas/GrooteW16}.
We shall prove in Section~3.2 that branching bisimulation quotients indeed preserve linearizability (Theorems 9 and 10).

These results provide us with a powerful tool for verifying linearizability, with several
advantages: (1) We can use existing  bisimulation checking tools (there are
many) to prove linearizability; (2) We can check linearizability on branching bisimulation
quotients, resulting in huge state space reductions; (3) Our approach does not
rely on prior identification of linearization points; (4) We can verify
progress properties in the same framework, using divergence-sensitive branching bisimulation.
Our approaches are summarized in Figure~\ref{fig:intro}.


To test the effectiveness  of our approaches, we have conducted a series of experiments
on more than a dozen concurrent data structures, using the existing proof toolbox
CADP~\cite{DBLP:journals/sttt/GaravelLMS13}, originally developed for concurrent systems.
The results of our experiments demonstrate that  huge state space reductions were
achieved due to quotient constructions. A new bug violating lock-freedom was found
and a known bug on linearizability was confirmed.


{\bf Organization}
Section 2 briefly reviews object systems and linearizability.
Section 3 introduces branching bisimulation   and  defines the quotient construction.
Section 4 presents our approach to checking progress properties.
Section 5 summarizes our experiments on various benchmarks.
Section 6 provides a  comparison with related work.  Section 7 concludes.

%% file: objsys.tex
\section{Object Systems and Linearizability}

\subsection{Object Systems}

The behaviors of a concurrent object can be adequately described as a labeled transition system.
We assume there is a language for describing concurrent algorithms, and
the language is equipped with an operational semantics to generate
labeled transition systems as defined below, also called ``object systems'',
from textual descriptions.
We will use the term ``object systems'' to refer to either the transition systems
or the program texts, depending on the context.

To generate an object's behaviour,
we use \emph{the most general clients}  \cite{Gotsman11,Liu13}, which
repeatedly invoke an object's methods in
any order and with all possible parameters.
We assume  a fixed collection $O$ of objects.

\begin{definition}[Labeled transition systems for concurrent objects] \label{lts}
A \emph{labled transition system} $\Delta$ is a quadruple $(S, \longrightarrow, {\cal A}, s_0)$
where
\begin{itemize}
  \item[$\bullet$] $S$ is the set of states,
 \item[$\bullet$]
${\cal A} = \setof{(t, \textsf{call}, \ $o$.m(n)), (t, \textsf{ret}(n'), \ $o$.{m}), (t,\tau)}{o \in O, t \in \{1 \ldots k\}}$, where $k$ is the number of threads,
is the set of actions.
  \item[$\bullet$] $\longrightarrow\ \subseteq S \times {\cal A} \times S$ is the transition relation,
   \item[$\bullet$] $s_0 \in S$ is the initial state.
\end{itemize}
\end{definition}
We shall write $s  \xrightarrow{a} s'$ to abbreviate $(s, a, s') \in \longrightarrow$.

When analysing the behaviours of a concurrent object, we are interested in
the interactions (i.e., call and return) between the object and its clients,
while the internal operations of the object are considered invisible.
Thus the visible actions of an object system are of the following
two forms:
$(t, \textsf{call}, o.m(n))$ and
$(t, \textsf{ret}(n'), o.{m})$,
where $t$ is a thread identifier. ${\footnotesize(t, \textsf{call}, o.m(n))}$
indicates an invocation of the method $m(n)$ of object $o$  by thread $t$
with the parameter $n$, and ${\footnotesize(t, \textsf{ret}(n'), o.m)}$ marks the
returning of a call to the method $m$ of $o$ by $t$
with the return value $n'$.
All other operations are regarded invisible and modeled by the silent
action $\tau$.

We write $s  \xrightarrow{\tau} s'$ to mean $s  \xrightarrow{(t,\tau)} s'$
for some $t$. 
A \emph{path} starting at a state $s$ of an object system
is a finite or infinite sequence $s \xlongrightarrow{a_1} s_{1} \xlongrightarrow{a_2} s_{2} \xlongrightarrow{a_3} \cdots$.
A {\em run} is a path starting from the initial state, which represents an entire computation
of the object system. A {\em trace} of state $s$ is a sequence of visible actions
obtained from a path of $s$ by omitting states and invisible actions,
which describes  the interactions of a client program with an object.

\subsection{Linearizability}

Linearizability is defined using \emph{histories}.
A history is a finite execution trace starting from the initial state and
consisting of call and return actions.
Given an object system $\Delta$,  its set of histories is denoted by ${\cal H}(\Delta)$.
If $H$ is a history and $t$ a thread, then the projection of $H$ on $t$, written $H|t$,
is called the subshitory of $H$ on $t$.
A history is \emph{sequential} if (1) it starts with a method call, (2) calls and returns alternate in the history, and (3)
each return matches immediately the preceding method call.
A sequential history is \emph{legal} if it respects the sequential specification of the object.
A call is \emph{pending} if it is not followed by a matching return.
Let $complete(H)$ denote the history obtained from $H$ by deleting all pending calls.

An operation $e$ in a history is a pair which consists of an invocation event $(t, \textsf{call}, o.m(n))$ and the matching response event $(t, \textsf{ret(n')}, o.m)$.
We shall use $e.call$ and $e.ret$ to denote, respectively, the invocation and
response events of an operation $e$.
The operation ordering in $H$ can be formally described using an irreflexive partial order $<_H$ by requiring that
$(e, e') \in\ <_H$ if $e.ret$ precedes $e'.call$ in $H$.
Operations that are not related by $<_H$ are said to be \emph{concurrent} (or overlapping).
If $H$ is sequential then $<_{H}$ is a total order.

The key idea behind linearizability is to compare concurrent histories to sequential histories.
We define the linearizability relation between histories.

\begin{definition}[Linearizability relation between histories]
$H \sqsubseteq_{\textsf{lin}} S$, read ``$H$ is linearizable \emph{w.r.t.} $S$'', if (1) $S$ is sequential, (2) $H|t = S|t$ for each thread $t$, and (3) $<_H~ \subseteq ~ <_{S}$.
\qed
\end{definition}

Thus $H \sqsubseteq_{\textsf{lin}} S$
if $S$ is a permutation of $H$ preserving (1) the order of actions in each thread, and (2) the non-overlapping method calls in $H$.
We use ${\cal H}(\Gamma)$ to denote the set of all histories of the sequential specification $\Gamma$.

\begin{definition}[Linearizability of object systems] \label{linearizability}
An object system $\Delta$ is \emph{linearizable w.r.t.\ a sequential specification} $\Gamma$, if $\forall H_1 \in {\cal H}(\Delta).$ $\, \left( \exists S \in {\cal H}(\Gamma). \, complete(H_1) \sqsubseteq_{\textsf{lin}} S \right)$.\qed
\end{definition}

An object is \emph{linearizable} if all its completed 
histories are linearizable \emph{w.r.t.} legal
sequential histories. Figure~\ref{fig:LinearExamples} shows
a linearizable history $H$ of a queue object
w.r.t. the legal sequential history $S$ and its thread subhistories.

\begin{figure}[htpb]
\vspace{-.8em}
\begin{center}
\scalebox{0.78}[0.6]{
\begin{tabular}{l|c||c|c|c}
1 & $(t_1, \textsf{call}, q.Enq(a))$ & $(t_1, \textsf{call}, q.Enq(a))$ & $(t_1, \textsf{call}, q.Enq(a))$ & $(t_2, \textsf{call}, q.Enq(b))$  \\
2 & $(t_2, \textsf{call}, q.Enq(b))$ & $(t_1, \textsf{ret()}, q.Enq)$   & $(t_1, \textsf{ret()}, q.Enq)$ & $(t_2, \textsf{ret()}, q.Enq)$  \\
3 & $(t_1, \textsf{ret()}, q.Enq)$   & $(t_2, \textsf{call}, q.Enq(b))$ & $(t_1, \textsf{call}, q.Deq)$ &  \\
4 & $(t_2, \textsf{ret()}, q.Enq)$   & $(t_2, \textsf{ret()}, q.Enq)$   & $(t_1, \textsf{ret(a)}, q.Deq)$&   \\
5 & $(t_1, \textsf{call}, q.Deq)$    & $(t_1, \textsf{call}, q.Deq)$    & &   \\
6 & $(t_1, \textsf{ret(a)}, q.Deq)$  & $(t_1, \textsf{ret(a)}, q.Deq)$  & &  \\
\hline
& $H$ & $S$ & $H\mid t_1$ & $H \mid t_2$
\end{tabular}
}
\end{center}
\vspace{-1.5em}
\caption{Example for a linearizable history and its thread subhistories.}
\label{fig:LinearExamples}
\end{figure}

Linearizability is a local property, i.e., a system is linearizable iff each object is linearizable.
Without loss of generality, we consider one object at a time.

\subsection{Linearizable Specification and Trace Refinement}

Given a concrete object system $\Delta$, we define its corresponding
{\em linearizable specification}~\cite{Gotsman11,Liu13,Feng},
denoted  $\Theta_{sp}$,
by turning the body of each
method in $\Delta$ into a single atomic block.
Such a specification allows non-terminating method calls which may overlap each other.
Thus, any method with non-terminating and overlapping execution intervals in the concrete implementation can be reproduced in the specification.
A method execution in a linearizable specification $\Theta_{sp}$ includes three main steps:
the call action ${\small(t, \textsf{call}, o.m(n))}$, the internal action $\tau$, and the return action ${\small(t, \textsf{ret}(n), o.m)}$.
The internal action corresponds to the computation based on the sequential specification of the object.
Each of the three actions is executed atomically.


Linearizability can be casted as trace refinement \cite{Hearn10,Liu13,Feng13}.
Trace refinement is a subset relationship between traces of two object systems,
an implementation 
and a specification.
Let $\mathit{trace}(\Delta)$ denote the set of all traces in $\Delta$.

\begin{definition}[Refinement]\label{refine}
Let $\Delta_1$ and $\Delta_2$ be two object systems.
$\Delta_1$ refines $\Delta_2$, written as $\Delta_1 \sqsubseteq_{tr} \Delta_2$,
if and only if $~\mathit{trace}(\Delta_1) \subseteq \mathit{trace}(\Delta_2)$.
\end{definition}

The following theorem shows that trace refinement exactly captures linearizability. A proof of this result can be found in \cite{Liu13}.

\begin{theorem} \label{lin-refine}
Let $\Delta$ be an object system and $\Theta_{sp}$ the corresponding specification.
All histories of $\Delta$ are linearizable
if and only if $\Delta \sqsubseteq_{tr} \Theta_{sp}$.
\end{theorem}

%% file: co-bis.tex
\section{Branching Bisimulation for Concurrent Objects}\label{branchbisim}

\subsection{Branching Bisimulation}

Branching bisimulation \cite{Glabbeek96} refines Milner's
weak bisimulation by requiring two related states should preserve
not only their own branching structure but also the branching
potentials of all intermediate states that are passed through.

\begin{definition}\label{co-sim}
Let ${\scriptsize \Delta = (S, \rightarrow, {\cal A}, s_0)}$ be an object system. A symmetric relation ${\cal R}$ on $S$ is a branching
bisimulation
if for all $(s_1, s_2) \in {\cal R}$ the following holds:
\begin{enumerate}
  \item if $s_1 \xlongrightarrow{a} s_1'$ where
  $a$ is a visible action,
  then there exists $s_2'$ such that
  $s_2 \xLongrightarrow{ } \xlongrightarrow{a} s_2'$
and $(s_1', s_2')\in \cal{R}$.

  \item if $s_1 \xlongrightarrow{\tau} s_1'$, then either ${(s_1', s_2)\in {\cal R}}$,
  or there exist $l$ and $s'_2$ such that
  $
  s_2 \xLongrightarrow{ } l \xlongrightarrow{\tau} s_2'
  $,
  $(s_1, l) \in {\cal R}$ and $(s_1', s_2')\in {\cal R}$.
\end{enumerate}
Let $\approx \DEF \bigcup \{ {\cal R} \mid {\cal R} \mbox{ is a branching bisimulation}\}$.
Then $\approx$ is the largest branching bisimulation and is
an equivalence relation.
\end{definition}

In the second clause of the above definition, for $s_2 \xLongrightarrow{ } l$
we only require $(s_1, l) \in {\cal R}$, without referring to the states that
are passed through in $s_2 \xLongrightarrow{ } l$. The following Stuttering Lemma, quoted from
\cite{Glabbeek96}, shows
that such omitting causes no problem.

\begin{lemma}\label{branching-property}
 If $r \xlongrightarrow{\tau} r_1
  \xlongrightarrow{\tau} \cdots \xlongrightarrow{\tau} r_m \xlongrightarrow{\tau} r'$ is a path such that $r \approx s$ and $r' \approx s$, then
  $r_i \approx s$ for all $i$ such that $1 \leq i \leq m$.

\end{lemma}

Thus the second clause in Definition~\ref{co-sim} can be expanded to:
\begin{enumerate}
  \item[{\it 2.}] if ${\footnotesize s_1 \xlongrightarrow{\tau} s_1'}$, then either ${(s_1', s_2)\in {\cal R}}$,
  or there exist $l_1, \cdots, l_i$, $i \geq 0$, and $s'_2$ such that
  $
  {\footnotesize s_2 \xlongrightarrow{\tau} l_1 \xlongrightarrow{\tau} \cdots \xlongrightarrow{\tau} l_i}{\footnotesize \xlongrightarrow{\tau} s_2'}
  $
  and
  ${\footnotesize (s_1, l_1) \in {\cal R}}, \cdots, {\footnotesize (s_1, l_i) \in {\cal R}}$, ${\footnotesize (s_1', s_2')\in {\cal R}}$.
\end{enumerate}

In contrast, branching potentials of the intermediate states are overlooked
in weak bisimulation~\cite{DBLP:books/daglib/0067019}.
As a result, weak bisimulation fails to preserve
linearization points.
An example showing this is deferred to  Appendix~\ref{append}.

For finite state systems, branching bisimulation can be computed in
polynomial time. The algorithm proposed in \cite{Groote90}
has time complexity
$O(|{\cal A}| + | S | \times | \longrightarrow |)$.
This result has recently been improved to
$O(|\longrightarrow| \times (log|Act| + log|S|))$ in
\cite{DBLP:conf/tacas/GrooteW16}.
\forget{
In particularly, the recent work \cite{DBLP:conf/tacas/GrooteW16} provides
a new algorithm to determine branching bisimulation in $O(|\longrightarrow| \times (log|Act| + log|S|))$,
which improves the existing algorithms such as \cite{Groote90} that has time complexity
$O(|{\cal A}| + | S | \times | \longrightarrow |)$.
}

\subsection{Checking Linearizability via Branching Bisimulation Quotienting}

Given an object system ${\footnotesize\Delta =(S, \xlongrightarrow{}, {\cal A}, s_0)}$, for any $s \in S$, let $[s]_{\approx}$ be the
equivalence class of $s$ under $\approx$, and
$S/\!\!\approx \  = \{[s]_{\approx} \mid s \!\in \!S\}$  the
set of the equivalence classes under $\approx$.

\begin{definition}[Quotient transition system] \label{quotient}
For an object system ${\footnotesize\Delta =(S, \xlongrightarrow{}, {\cal A}, s_0)}$,
the quotient transition system $\Delta/ {\approx}$ is defined as:
${\scriptsize
   \Delta / {\approx} = (S/ {\approx}, \xlongrightarrow{}_{\approx}, Act,  [s_0]_{\approx})
}$,
where the transition relation
$\xlongrightarrow{}_{\approx}$ is generated by the following rules:
  $$
  {\footnotesize
  \begin{array}{ll}
  (1) \frac{\displaystyle s \xlongrightarrow{\alpha} s' }
  {\displaystyle [s]_{\approx} \xlongrightarrow{\alpha}_{\approx} [s']_{\approx}} \ (\alpha \neq \tau) ~~~~
  (2) \frac{\displaystyle s \xlongrightarrow{\tau} s' }
  {\displaystyle [s]_{\approx} \xlongrightarrow{\tau}_{\approx} [s']_{\approx}} \ ((s, s') \not\in \approx)
  \end{array}
  }
  $$
\end{definition}

\begin{theorem} \label{co-lin}
$\Delta /{\approx}$ preserves linearizability. That is,
$\Delta$ is linearizable
if and only if $\Delta /{\approx}$ is linearizable.
\end{theorem}

\begin{Proof}
Let $\Theta_{sp}$ be the corresponding specification of $\Delta$.
Then it is also the corresponding specification of $\Delta/{\approx}$.
From Definition \ref{co-sim}, it is easy to see that ${\small \mathit{trace}(\Delta) =}$ ${\small \mathit{trace}(\Delta/{\approx})}$.
Thus, we have ${\small\mathit{trace}(\Delta) \subseteq \mathit{trace}(\Theta_{sp})}$
iff  ${\small\mathit{trace}(\Delta /{\approx} )}$ ${\small\subseteq \mathit{trace}(\Theta_{sp})}$.
By Definition \ref{refine}, ${\small \Delta \sqsubseteq_{tr} \Theta_{sp}}$ iff
${\small\Delta /{\approx} \sqsubseteq_{tr} \Theta_{sp}}$.
Further, by Theorem  \ref{lin-refine}, it follows that $\Delta$ is
linearizable w.r.t. $\Theta_{sp}$ iff $\Delta /{\approx}$ is linearizable w.r.t. $\Theta_{sp}$.
\qed
\end{Proof}

\begin{theorem} \label{co-quo-lin}
An object system $\Delta$ with the corresponding specification $\Theta_{sp}$
is linearizable
if and only if $\Delta / {\approx} ~\sqsubseteq_{tr} ~\Theta_{sp} / {\approx}$.
\end{theorem}

\begin{Proof}
By Theorems \ref{lin-refine} and \ref{co-lin}.
\qed
\end{Proof}

It is well-known that deciding trace inclusion is PSPACE-complete.
Hence verifying linearizability in an automated manner by directly resorting to
Definition~\ref{linearizability}
is infeasible in practice.
Since an object system  contains a lot of invisible transitions,
among them only a few are responsible for changing the system's states,
and non-blocking
synchronization usually  generate a large number of interleavings,
its branching bisimulation quotient is usually much smaller than
the object system itself. Furthermore, branching bisimulation quotients can be
computed efficiently. Thus
Theorem~\ref{co-quo-lin} provides us with a practical solution
to the linearizability verification problem:
\begin{quote}
Given an object system
$\Delta$ and a specification $\Theta_{sp}$,
first compute their branching bisimulation quotients
$\Delta / {\approx}$ and $\Theta_{sp} / {\approx}$, then check
$\Delta / {\approx} ~\sqsubseteq_{tr} ~\Theta_{sp} / {\approx}$.
\end{quote}

In practice, this approach results in huge reductions of state spaces.
Details of our experiments  are reported in Section~\ref{sec:exp}.

\forget{

\subsection{A Discussion on Weak Bisimulation}

We now take Michael-Scott lock-free queue (MS queue)~\cite{DBLP:conf/podc/MichaelS96} as a realistic example,
which is used in \textsf{java.util.concurrent} package, to show that branching bisimulation
is an important notion on semantics equivalence of states to capture the computation effect for the concurrent objects.
The concrete algorithm of the queue is shown in Figure~\ref{pic:MS}. The queue is implemented by a linked-list, where \texttt{Head} and \texttt{Tail} refer to the first and the last node respectively. There are two LPs in method \texttt{deq()}:
at \atl{28} when $\mathtt{cas(Head, h, s)}$ succeeds,
and at \atl{20}, a \emph{non-fixed LP} that depends on the future executions.
The intuition is as follows.
If we read \emph{null} from \texttt{h.next} at \atl{20}, but interleavings with other threads 
before \atl{21} yield a change of \texttt{Head} such that the condition on \texttt{Head} at \atl{21} fails, then the method has to restart the loop, and \atl{20} may not be the LP.

\begin{figure}
 \begin{minipage}{.46\textwidth}
 \vspace{0ex}
 \begin{lstlisting}[basicstyle=\scriptsize\ttfamily, escapechar=|]
 L01 enq(v) {
 L02  local x,t,s,b;
 L03  x:=new_node(v);
 L04  while(true) {
 L05   t:=Tail; s:=t.next;
 L06   if (t=Tail) {
 L07    if (s=null) {
 L08    b:=cas(&(t.next),s,x);
 L09     if (b) {
 L10      cas(&Tail,t,x);
 L11      return true; }
 L12    }else cas(&Tail,t,s);
 L13   }
 L14  }
 L15 }

 L16 deq() {
 L17  local h,t,s,v,b;
 L18  while(true) {
 L19   h:=Head; t:=Tail;
 L20   |\setlength{\fboxsep}{1pt}\lcolorbox{yellow}{s:=h.next;}|
 L21    |\setlength{\fboxsep}{1pt}\lcolorbox{yellow}{if (h=Head)}|
 L22    if (h=t) {
 L23      if (s=null)
 L24         return EMPTY;
 L25      cas(&Tail,t,s);
 L26    }else {
 L27      v :=s.val;
 L28      |\setlength{\fboxsep}{1pt}\lcolorbox{yellow}{b:=cas(\&Head,h,s);}|
 L29      if(b) return v;}
 L30  } }
\end{lstlisting}
\end{minipage}
\begin{minipage}{0.4\textwidth}
\vspace{-3.5ex}
\includegraphics[scale=.58]{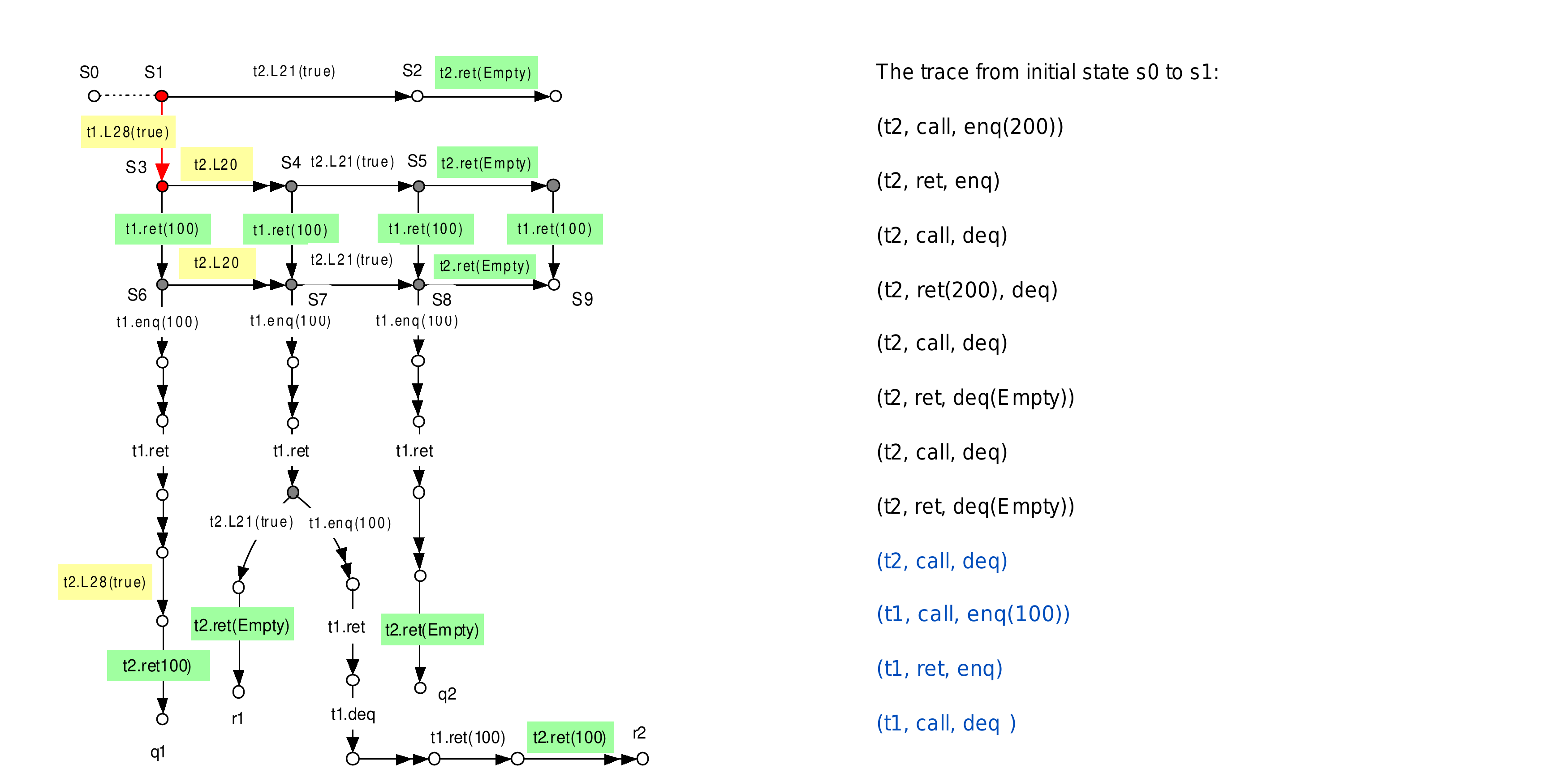}
\end{minipage}
 \caption{MS lock-free queue and its (part) transition system.} \label{pic:MS} \vspace{-4.3ex}
 \end{figure}


We assume 2 threads and each invokes methods enqueue and dequeue for 5 times.
Its part transition system is shown in Figure~\ref{pic:MS}, where
$s_0$ is the initial state. The omitted history from $s_0$ to $s$ is:
(${\small(t_2,\texttt{call,enq(200))}}$,
 ${\small(t_2,\texttt{ret,enq})}$,
 ${\small(t_2,\texttt{call,deq})}$,
 ${\small(t_2,\texttt{ret(200),deq})}$,
 ${\small(t_2,\texttt{call,deq})}$,
 ${\small(t_2,\texttt{ret(Empty), deq})}$,
 ${\small(t_2,\texttt{call, deq})}$,
 ${\small(t_2,\texttt{ret(Empty), deq})}$,
 ${\small(t_2,\texttt{call, deq})}$,
 ${\small(t_1,\texttt{call, enq(100)})}$,
 ${\small(t_1,\texttt{ret, enq})}$).
At state $s$, the current queue is non-empty with element 100,
and thread $t_2$ has invoked 5 times, where the last invocation ${\small(t_2, \texttt{call, deq})}$ reads $\texttt{Tail}$ and $\texttt{Head}$ at L19
before the last invocation ${\small(t_1,\texttt{call,enq(100)})}$. Thus,
at state $s$, we have $s \xlongrightarrow{t_2.L20} \xlongrightarrow{t_1.deq} s_1$.
Figure~\ref{pic:MS} shows the different subsequent traces of $s_1$.

\begin{itemize}
  \item For the execution from $s_1$ to $s_2$, it is obvious that $t_2$ always returns ${\small t_2.\texttt{ret(Empty)}}$;
  \item For the execution from $s_1$ to $s_5$, since step $s_1 \xlongrightarrow{t_1.L28} s_3$ updates $\texttt{Head}$ successfully,
        which make the later check of $t_2$ at $L21$ fail and restart. Since the queue is empty for the path from $s_3$ to $s_5$,  $t_2$ returns ${\small t_2.\texttt{ret(Empty)}}$;
  \item For the execution from $s_3$ to $q_1$, $t_1$ updates $\texttt{Head}$ at step $s_1 \xlongrightarrow{t_1.L28} s_3$, and then enqueues 100 successfully in the path from $s_3$ to $s_6$. Thus at $s_6$, $t_2$ check $L21$ fail and restart. In the new round execution, $t_2$ returns ${\small t_2.\texttt{ret(100)}}$;
  \item For the execution from $s_4$ to $r_1$, since $t_2$ has executed $L20$ from $s_3$ to $s_4$ and reads the updated $\texttt{Head}$ at $s_3$.
        Thus although $t_1$ enqueues 100 successfully in the path from $s_4$ to $s_7$, $t_2$ checks L21 still as true. Therefore, $t_2$ returns ${\small t_2.\texttt{ret(Empty)}}$;
  \item For the execution from $s_4$ to $r_2$, $t_1$ enqueues 100 twice, and then $t_1$ dequeues 100 that updates $\texttt{Head}$.
        This makes the check $h= Head$ at L21 by $t_2$ fail. When $t_2$ restarts, it returns ${\small t_2.\texttt{ret(100)}}$ at $r_2$.
  \item For the execution from $s_5$ to $q_2$, it is obvious that $t_2$ returns ${\small t_2.\texttt{ret(Empty)}}$.
\end{itemize}

From the above executions, since the path from $s_3$ to $q_1$
and the path from $s_4$ to $r_1$ which are labeled with the same sequence of visible actions,
but generate different return values ${\small t_2.\texttt{ret(100)}}$ and ${\small t_2.\texttt{ret(Empty)}}$ respectively, we have $s_3 \not\approx s_4$.
Similarly, $s_4 \not\approx s_5$. Further, since ${\small(t_2,\texttt{call, deq})}$ is the last invocation that
always returns ${\small t_2.\texttt{ret(Empty)}}$, it is easy to see $s_2 \approx s_5$.
For the step $s_1 \xlongrightarrow{\tau} s_2$ and the $\tau$-path $s_3 \xlongrightarrow{\tau} s_4 \xlongrightarrow{\tau} s_5$,
branching bisimulation requires to check each intermediate state along the $\tau$-path. Because $s_3 \not\approx s_4 \not\approx s_5$,
it follows $s_1 \not \approx s_3$.
However, observation equivalence directly compares
the last states $s_2$ and $s_5$, and does not need to consider the computation effect of intermediate state $s_4$.
Because $s_2$ and $s_5$ are weak bisimilar, it follows that $s_1$ and $s_3$ are weak bisimilar.

However $s_1 \xlongrightarrow{t_1.L28} s_3$ is a linearization point for successful dequeue.
Therefore, observation equivalence cannot be as a sufficient notion to capture linearization points.

\textbf{linearization points.}
To prove linearizability, the most common approach is to
identify a time point in the code of each method
and show that the entire effect of the method takes place at this point instantaneously.
This time instant is called \emph{linearization points} of that method.
Within each method execution, exactly one linearization point must occur.

In accordance with the object system,
a method may take a sequence of $\tau$-steps, while synchronizing with
other methods to complete a call.
If the object system is linearlizable, then among these $\tau$-steps,
there is only one step that takes effect to change the object's state (or a pure execution to complete the method call)
while the other $\tau$-steps do not.
The $\tau$-step which takes effect to complete the method call corresponds to a linearization point.

An object system may have many linearization point. Informally,
A lineairizaiton point is \emph{static}, if it can be statically located in the source code
such that every time the underling code is executed, it is a linearization point.
Otherwise, it is a \emph{non-fixed} linearization point.
Non-fixed linearization points are complex to be fixed in the source code
since these linearization points are conditional, which depends on the interleavings
with other threads, and some of which are even not inside the source code of its own method call.
As presented, for method deq in MS queue, the linearization point (L20) for the empty case is non-fixed
only if (L21) is true and later executes return EMPTY.

}

%% file: divergence.tex
\section{Progress Properties}
We exploit \emph{divergence-sensitive} branching bisimulation between a concrete and 
an abstract object to verify progress properties
of concurrent objects. The main result that we will establish is that for 
divergence-sensitive branching bisimilar abstract and concrete objects, 
it suffices to check progress properties on the abstract objects.

Lock-freedom and wait-freedom are the most commonly used progress properties in non-blocking concurrency~\cite{Herlihy08}.
Informally, a method is \emph{wait-free} if it satisfies that each thread finishes a method call in
a finite number of steps, while lock-freedom guarantees that some thread can complete a started method call in
a finite number of steps~\cite{Herlihy08}.
Their formal definitions specified using next-free LTL are given in~\cite{Petrank09,icfem06}.

A linearizable specification is an atomic abstraction of concurrent objects.
It is not hard to see that the object system for the linearizable specification satisfies the lock-free property.
To obtain wait-free object systems, we need to enforce some fairness assumption on transition systems to guarantee the fair scheduling of processes.
The most common fairness properties (such as strong and weak fairness) can all be expressed in next-free LTL.

\begin{lemma} \label{spec-lockfree}
The linearizable specification $\Theta_{sp}$ is lock-free.
\end{lemma}

\begin{Proof}
$\Theta_{sp}$ consists of a single atomic block (see Section 2.3), of which the internal execution corresponds to the computation of the sequential specification that
by assumption is always safe and terminating.
Hence for any run of $\Theta_{sp}$, there always exists one thread to complete its method call in finite number of steps. \qed
\end{Proof}

A pending call of a run is {\em blocking} if it requires to wait for other method call to complete.
Let us recall the Herlihy and Wing queue. When the queue is empty, 
the call of Deq is blocking, as it will stay forever in a $\tau$-loop (e.g., $s \xlongrightarrow{\tau} s_1 \dar s$ in Figure~\ref{HW-traces})
that does not perform any return action
if no element is enqueued.
Such behavior 
is called \emph{divergent}.
To distinguish infinite series of internal transitions from finite ones, we treat divergence-sensitive branching bisimulation \cite{Glabbeek96}.

\begin{definition}[Divergence sensitivity] \label{def-div}
Let $\Delta = (S, \longrightarrow, {\cal A}, s_0)$ be an object system and ${\cal R}$ an equivalence relation on $S$.
\begin{itemize}
\item
A state $s \in S$ is ${\cal R}$-divergent if there exists an infinite path $s \xlongrightarrow{a_1} s_1 \xlongrightarrow{a_2} s_2 \xlongrightarrow{} \cdots$ such that $(s, s_j) \in {\cal R}$ for all $j >0$.
\item
${\cal R}$ is divergence-sensitive if for all $(s_1, s_2) \in {\cal R}$: $s_1$ is divergent iff $s_2$ is divergent.
\end{itemize}
\end{definition}
\begin{definition}[\cite{Glabbeek96}]\label{thm-diver}
States $s_1, s_2$ in object system $\Delta$ are divergent-sensitive branching bisimilar, denoted $s_1 \approx_{div} s_2$, if there exists a divergence-sensitive branching bisimulation $\cal R$ on $\Delta$ such that $(s_1, s_2) \in {\cal R}$.
\end{definition}

This notion is lifted to object systems in the standard manner, i.e., object systems $\Delta_1$ and $\Delta_2$ are divergent-sensitive branching bisimilar whenever their initial states are related by $\approx_{div}$ in the disjoint union of $\Delta_1$ and $\Delta_2$.

Divergence-sensitive branching bisimulation implies (next-free) LTL and CTL$^*$-equiva\-len\-ce \cite{Groote90}.
This also holds for countably infinite transition systems that are finitely branching.
Thus, $O \approx_{div} \Theta$ implies the preservation of all next-free LTL and CTL$^*$-formulas.
Since the lock-freedom (and other progress properties~\cite{icfem06}) can be formulated in next-free LTL,
for abstract object $\Theta$ and concrete object $O$, it can be preserved by the relation $O \approx_{div} \Theta$.

For a concrete object its abstract object is a coarser-grained concurrent implementation.
If an appropriate abstract object for a concrete algorithm can be provided, one can check progress properties on the (usually much simpler) abstract objects.
For finite-state abstract programs, off-the-shelf model checking tools can be readily applied to check their properties.

\begin{theorem} \label{thm-dynamic-lp}
For the abstract object $\Theta$ and concrete object $O$, if $O \approx_{div} \Theta$, then $\Theta$ is lock-free iff $O$ is lock-free.
\end{theorem}

The process of constructing an abstract object is often manually and the discussion about it is outside the scope of the paper.
However, for objects with \emph{static linearization points} such as Treiber stack~\cite{Treiber} and stacks with hazard pointers~\cite{Michael04},
since there is only one linearization point for each method,
which behaves in accordance with the behaviour of atomic block of the linearizable specification,
the specification can be directly as the abstract object. Thus,
we can provide an easier way to verify linearizability and lock-free property together for this kind of object.

\begin{corollary} \label{thm-static-lp}
Let $O$ be an object with static linearization points and $\Theta_{sp}$ its specification.
If ${\small O \approx_{div} \Theta_{sp}}$, then $O$ is lock-free and linearizable.
\end{corollary}

\begin{Proof}
For lock-free property, it is straightforward by Lemma~\ref{spec-lockfree} and Theorem~\ref{thm-dynamic-lp}.
For linearizability, since ${\small O \approx_{div} \Theta_{sp}}$, it follows ${\small \mathit{trace}(O) =}$ ${\small \mathit{trace}(\Theta_{sp})}$.
By Definition \ref{refine}, ${\small O \sqsubseteq_{tr} \Theta_{sp}}$.
Thus, by Theorem  \ref{lin-refine}, $O$ is linearizable.
\qed
\end{Proof}

%% file: experiments.tex
\section{Experiments}\label{sec:exp}

To illustrate the effectiveness and efficiency of our techniques for proving linearizability as well as progress properties,  we conduct experiments on
a number of practical concurrent algorithms,
including 4 queues (3 lock-free, 1 lock-based), 4 lists (1 lock-free, 3 lock-based), 3 (lock-free) stacks and 2 extended CAS (compare-and-swap) operations,
some of which are used in the \textsf{java.util.concurrent} package.
We employ the Construction and Analysis of Distributed Processes (CADP)~\cite{DBLP:journals/sttt/GaravelLMS13} toolbox\footnote{http://cadp.inria.fr/} for these experiments.
The case studies are summarized in Table~\ref{tab:exov}.
\forget{
Our experiments found a new bug violating lock-freedom in the revised Treiber stack~\cite{DBLP:conf/concur/FuLFSZ10} and confirmed a known bug in the HM lock-free list~\cite{Herlihy08}.
These bugs were found using a fully automated verification procedure on finite-state instances of the concurrent data structures.
}

\begin{table}[htbp]\caption{Case studies and overview of their verification.}\label{tab:exov}
\centering
\scalebox{0.72}[0.63]{
\begin{tabular}{|l|c|c|c|c|}
\hline
\hline
 \qquad {\footnotesize Case study } & {\footnotesize Linearizability \& Lock-freedom} & {\footnotesize Non-fixed LPs} &{\footnotesize branch bisim./trace ref.}& {\footnotesize Java Pkg} \\
\hline
{\footnotesize 1. Treiber stack~\cite{Treiber} } & $\checkmark$ &   & $\checkmark$ &  \\
\hline
{\footnotesize 2. Treiber stack+HP~\cite{Michael04} } & $\checkmark$ &   &$\checkmark$ &  \\
\hline
{\footnotesize 3. Treiber stack+HP~\cite{DBLP:conf/concur/FuLFSZ10} } & $\pmb{\times}$ {\footnotesize Lock-freedom} &   & $\pmb{\times}$ &  \\
\hline
{\footnotesize 4. MS queue~\cite{DBLP:conf/podc/MichaelS96} } & $\checkmark$ &  $\checkmark$ & $\checkmark$ & $\checkmark$ \\
\hline
{\footnotesize 5. DGLM queue~\cite{DBLP:conf/forte/DohertyGLM04} } & $\checkmark$ & $\checkmark$  & $\checkmark$ &  \\
\hline
{\footnotesize 6. CCAS~\cite{DBLP:conf/popl/TuronTABD13} } & $\checkmark$ & $\checkmark$ & $\checkmark$ &  \\
\hline
{\footnotesize 7. RDCSS~\cite{DBLP:conf/wdag/HarrisFP02} } & $\checkmark$ & $\checkmark$  & $\checkmark$ &  \\
\hline
\hline
\qquad {\footnotesize Case study } & {\footnotesize Linearizability} & {\footnotesize Non-fixed LPs} &{\footnotesize branch bisim./trace ref.}& {\footnotesize Java Pkg} \\
\hline
{\footnotesize 8. Fine-grained syn. list~\cite{Herlihy08} } & $\checkmark$ &   & $\checkmark$ &  \\
\hline
{\footnotesize 9-1. HM lock-free list~\cite{Herlihy08}} & $\pmb{\times}$ {\footnotesize Linearizability} &  $\checkmark$  &$\pmb{\times}$ & \\
\hline
{\footnotesize 9-2. HM lock-free list (revised) } & $\checkmark$ &  $\checkmark$  & $\checkmark$ &  $\checkmark$  \\
\hline
{\footnotesize 10. Optimistic list~\cite{Herlihy08} } & $\checkmark$ & $\checkmark$   & $\checkmark$ &  \\
\hline
{\footnotesize 11. Heller \emph{et al.} lazy list~\cite{DBLP:journals/ppl/HellerHLMSS07} } & $\checkmark$ &   & $\checkmark$ &  \\
\hline
{\footnotesize 12. MS two-lock queue~\cite{DBLP:conf/podc/MichaelS96}} & $\checkmark$ &   & $\checkmark$ &  \\
\hline
{\footnotesize 13. Herlihy-Wing queue~\cite{Herlihy90} } & $\checkmark$ & $\checkmark$  & $\checkmark$ &  \\
\hline
\end{tabular}}
\vspace{.6ex}
\vspace{-1ex}
\end{table}

\subsection{Proving linearizability and progress properties}

Linearizability has been proven by checking trace refinement between two branching bisimilar quotients---the concrete object and its specification, cf.\ Figure~\ref{fig:intro} (a) and Theorem~\ref{co-quo-lin}. Our technique does not rely on linearization points and can check all algorithms covered in~\cite{Feng13}.
As indicated in Table~\ref{tab:exov}, all but one data structure in the
case study are linearizable.

Progress properties were checked by checking divergence-sensitive branching bisimilarity between an abstract and concrete object, cf.\ Figure~\ref{fig:intro} (b) and Theorem~\ref{thm-dynamic-lp} and Corollary~\ref{thm-static-lp}.
We successfully verified lock-freedom for 6 algorithms. For objects with non-fixed linearization points,
abstract objects were constructed for MS queue, DGLM queue, CCAS and RDCSS.
For objects with static linearization points, no abstract objects need to be built (see Corollary~\ref{thm-static-lp}).
Our technique can verify lock-freedom of complex algorithms that are not included in~\cite{Feng14},
such as CCAS, RDCSS and the Trebier stack with hazard pointers (a garbage collection mechanism).
The details of verification results can be found in~\cite{appendix}.

\subsection{Automated bug hunting}

Our techniques are fully automated (for finite-state systems) and rely on efficient existing algorithms.
In contrast to proof techniques~\cite{Victor08,Victor10,Feng13,Feng14,Colvin06} for linearizabilty and progress,
our approach is able to generate counterexamples in an automated manner.
As indicated in Table~\ref{tab:exov}, we found a single linearizability violation and a lock-freedom violation.
\begin{enumerate}
\item
We found a---to our knowledge so far unknown-violation of lock-freedom in the revised Treiber stack~\cite{DBLP:conf/concur/FuLFSZ10}.
This revised version avoids the ABA problem at the expense of violating the wait-free property of hazard pointers in the original algorithm~\cite{Michael04}.
We found this bug by an automatically generated counterexample of divergence-sensitive branching bisimilarity checking by CADP with just two concurrent threads.
The error-path ends in a self-loop in which one thread keeps reading the same hazard pointer value of another thread
without making any progress.
\item
Our experiments confirmed a (known) bug in the HM lock-free list~\cite{Herlihy08} which was amended in the online errata of~\cite{Herlihy08}.
The counterexample is generated by the trace inclusion checking on the quotients of the concrete versus the specification.
It consecutively removes the same item twice, which violates the specification of being a list.
\end{enumerate}

\subsection{Efficiency and state-space savings}

Checking branching bisimilarity as well as computing  branching bisimulation quotients are  efficient; they both can be done in polynomial time.
This stands in contrast to directly checking trace refinement---the main technique so far for model checking linearizability---which is PSPACE-complete.
The result of our experiments show that checking lock-freedom and linearizability for models with millions of states is practically feasible.

\begin{figure}[htpb]
\centering
\begin{narrow}{-.0ex}{-.0ex}
\includegraphics[scale=.178]{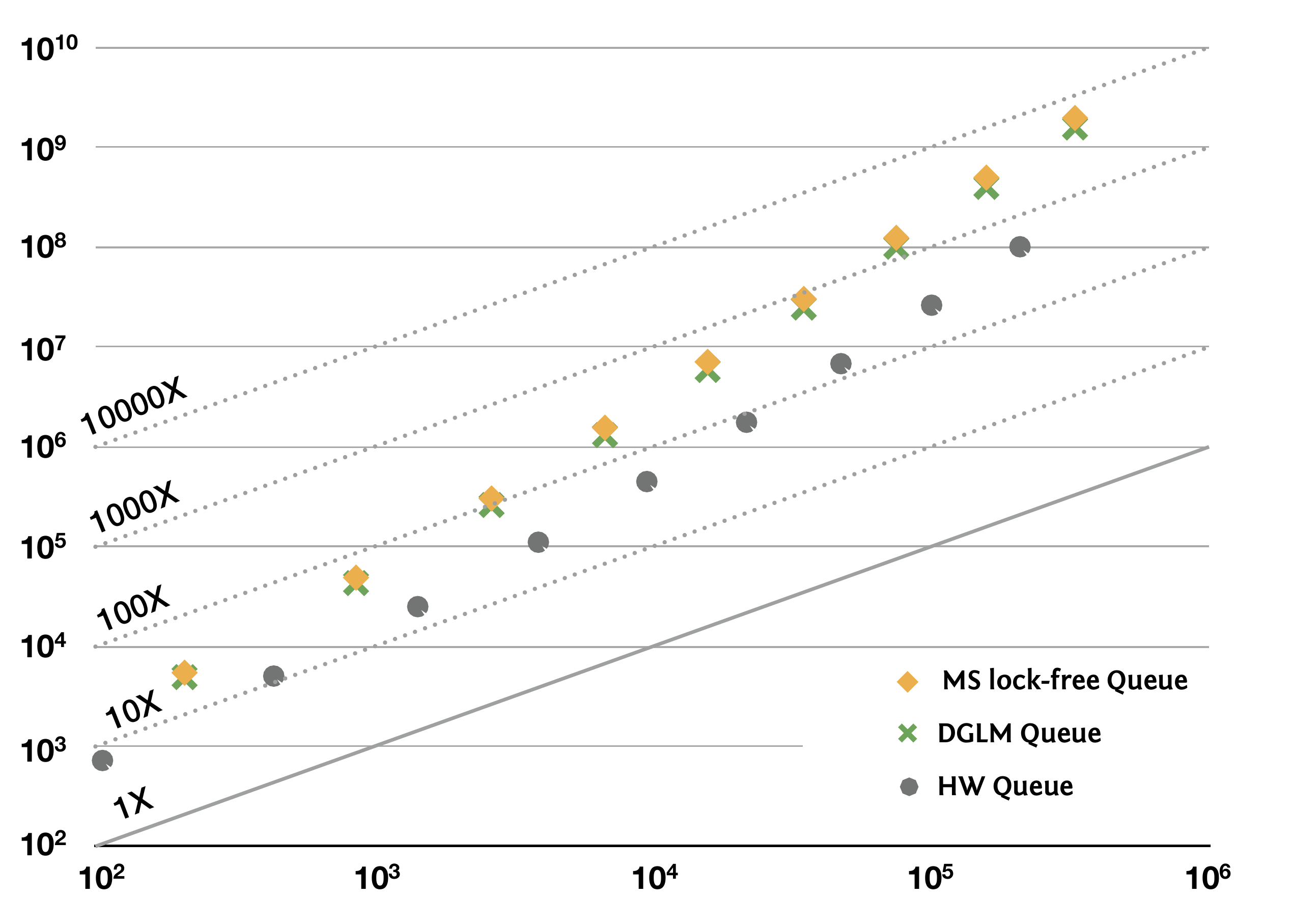}\vspace{-1ex}
\includegraphics[scale=.178]{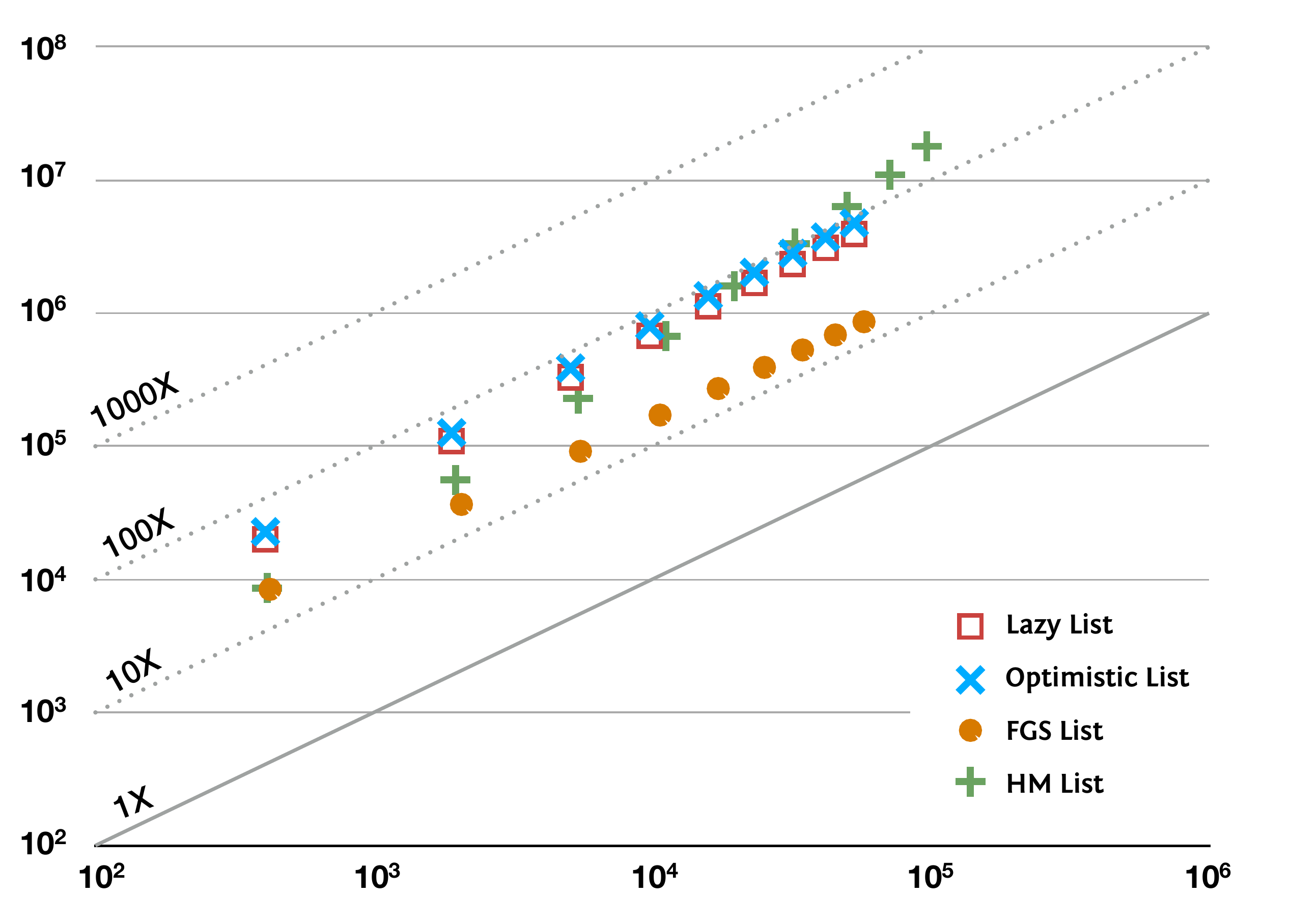}\vspace{-.5ex}
\includegraphics[scale=.178]{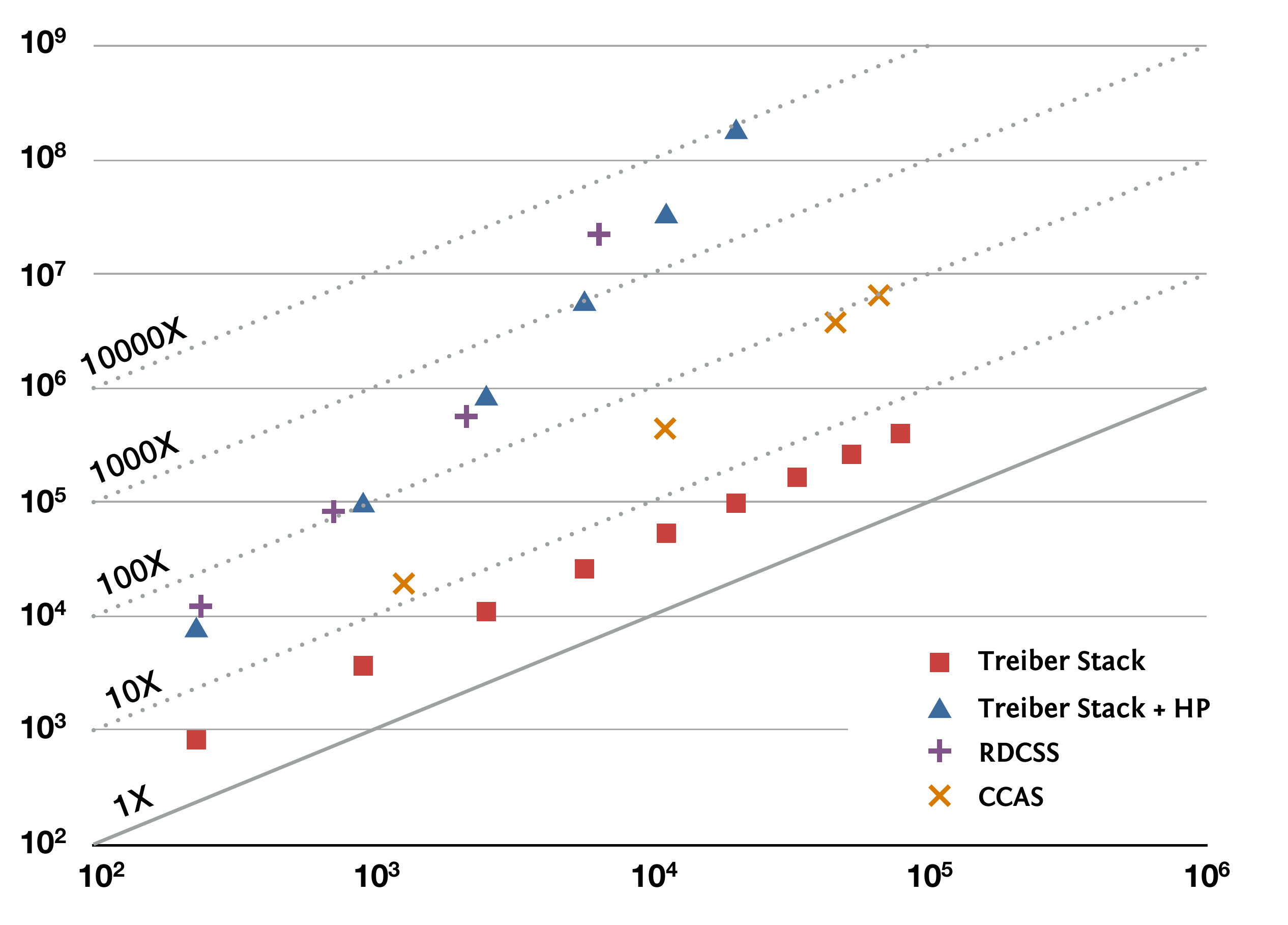}
\end{narrow}
\caption{State-space reduction using $\approx$-quotienting.}\label{fig:total}\vspace{-1ex}
\end{figure}

All experiments run on a server which is equipped with a 4$\times$12-core AMD CPU @ $2.1$ GHz and $192$ GB memory under 64-bit Debian 7.6.
Figure~\ref{fig:total} shows the state-space savings for 11 algorithms (for two threads invoking methods for 2-10 times).
Note that both the $x$- and the $y$-axis are in log-scale; for the sake of clarity we have indicated the lines with state space reduction factor 1 up to 10000 explicitly.
Branching bisimulation quotient construction has  yielded state-space savings of up to four orders of magnitude in the best cases, and to two to three orders for most cases.
And in general, for the non-blocking implementation, the larger the system the higher the state space reduction factor.
The largest reductions were obtained for the Treiber stack with hazard pointers (Treiber stack+HP) and the MS lock-free queue yielding a quotient with 0.01\% and 0.02\% of the size of the concrete objects, respectively. Verifying linearizability directly on the concrete state space would be practically infeasible.

%% file: relatedwork.tex
 \section{Related Work}
Linearizability has been intensively investigated in the literature.
A comparison with all works goes outside the scope of this paper; instead, we focus on the closest related works.

A plethora of proof-based techniques has been developed for verifying linearizability.
Most are based on rely-guarantee reasoning~\cite{Victor08,Victor10,Feng13}, or establishing simulation relations~\cite{Colvin06,Derrick11,Schellhorn12}.
These techniques often involve identifying linearization points which is a manual non-trivial task.
Of the more recent works, Liang \emph{et al.}~\cite{Feng13} propose a program logic tailored to rely-guarantee reasoning to verify complex algorithms.
This method is applicable to a wide range of popular non-blocking algorithms but is restricted to certain types of linearization points.
Challenging algorithms such as the Herlihy-Wing queue (\cite{Herlihy90} and \cite{DBLP:conf/popl/DoddsHK15}) fall outside this method.
Our techniques do not rely on identifying linearization points, and are aimed to exploit established notions from concurrency theory.

Model checking methods to verify linearizability have been proposed in e.g., \cite{Liu13,Alur10,spin09,pldi10}.
Liu et al.~\cite{Liu13} formalize linearizability as trace refinement and use partial-order and symmetry reduction techniques to alleviate the state explosion problem.
Their experiments are limited to simple concurrent data structures such as counters and registers, and their technique is not applicable to checking progress properties.
\forget{
not the classic algorithms in our work, their state-space savings are smaller than obtained using branching bisimulation in this paper (up to a factor $10^3$ to $10^4$), and their technique is not applicable to checking progress.

(Since the branching bisimulation quotients only contain linearization points---that are essential to linearizability---as well as the required call and returns, we feel that no further reduction can be obtained without sacrificing the preservation of linearizability and progress properties.)
}
Cerny \emph{et al.}~\cite{Alur10} propose method automata to verify linearizability of concurrent linked-list implementations, which
is restricted to two concurrent threads.
An experience report with the model checker SPIN~\cite{spin09} introduces an
automated procedure for verifying linearizability, but the method relies on manually annotated linearization points.


For the verification of progress properties, \cite{Gotsman11,Feng14,Feng} recently propose refinement techniques with termination preservation.
These techniques are limited to checking lock-freedom of some non-blocking algorithms (e.g., Treiber stack, MS and DGLM queues).
Neither more complex non-blocking algorithms nor other progress properties are discussed.
Our approach can check a large class of progress properties---in fact all properties expressible in CTL$^*$ (containing LTL) without next.
Our experiments treat 7 non-blocking algorithms and found a lock-free property violation in the revised stack~\cite{DBLP:conf/concur/FuLFSZ10}.
Some formulations of progress properties using next-free LTL are discussed in~\cite{Petrank09,icfem06}.

\section{Conclusion}

This paper proposed to exploit branching bisimulation (denoted $\approx$) --- a well-established notion in the field of concurrency theory --- for proving linearizability and progress properties of concurrent data structures.
A concurrent object $O$ is linearizable w.r.t.\ a linearizable specification $\Theta_{sp}$ iff their quotients under $\approx$ are in a trace refinement relation.
Unlike competitive techniques, this result is independent of the type of linearization points.
If the abstract and concrete object are divergence-sensitive branching bisimilar, then progress properties of the --- typically much smaller and simpler --- abstract object carry over to the concrete object.
This entails that progress properties such as lock- and wait-freedom (in fact all progress properties that can be expressed in the next-free fragment of CTL$^*$) can be checked on the abstract program.
Our approaches can be fully automated for finite-state systems.
We have conducted experiments on 13 popular concurrent data structures yielding promising results.
In particular, the fact that counterexamples can be obtained in an automated manner is believed to be a useful asset.
Our experiments confirmed a known linearizability bug and revealed a new lock-free property violation.

%% file: appendix.tex
\appendix

\section{A Discussion on Weak Bisimulation}\label{append}

Weak bisimulation, $\approx_w$, is obtained by replacing the second clause of
Definition~\ref{co-sim} with:

\begin{enumerate}
  \item[{\it 2.}] if $s_1 \xlongrightarrow{\tau} s_1'$, then
either ${(s_1', s_2)\in {\cal R}}$,
  or there exists $s'_2$ such that
  $
  s_2 \xLongrightarrow{ } \xlongrightarrow{\tau} s_2'
  $
  and $(s_1', s_2')\in {\cal R}$.
\end{enumerate}

Compared with branching bisimulation, weak bisimulation does not require the intermediate
states passed through to be matched. We present an example  showing that,
because of this, weak bisimulation failed to preserve linearization points.

The example is
Michael-Scott lock-free queue (MS queue)~\cite{DBLP:conf/podc/MichaelS96},
shown in Figure~\ref{pic:MS}.
The queue is implemented by a linked-list, where \texttt{Head} and \texttt{Tail} refer
to the first and the last node respectively.
It provides two methods: (1) \texttt{enq(v)}, which inserts an element in the end of the
queue; and (2) \texttt{deq}, which removes
the first element in the queue if there is one, and returns EMPTY otherwise.

\vspace{-.2em}
\begin{figure}[htpb]\centering
\begin{narrow}{0ex}{-1ex}
{\small
\begin{minipage}{.45\textwidth}
\begin{lstlisting}[basicstyle=\scriptsize\ttfamily,mathescape,escapechar=|]
1 enq(v) {	
2 local x, t, s, b;		
3 x := cons(v, null);	
4 while (true) {		
5  t := Tail; s := t.next;
6  if (t = Tail) {		
7   if (s = null) {	
8     |\setlength{\fboxsep}{1pt}\lcolorbox{yellow}{b:=cas(\&(t.next),s,x);}|
9    if (b) {		
10    cas(&Tail,t,x);	
11    return; }			
12  }else cas(&Tail, t,s);	
13 }				
14$\;$}
15}
\end{lstlisting}
\end{minipage}
\begin{minipage}{.47\textwidth}
\begin{lstlisting}[basicstyle=\scriptsize\ttfamily,escapechar=|]
16 deq() {
17   local h, t, s, v, b;
18   while (true) {
19    h := Head; t := Tail;
20    |\setlength{\fboxsep}{1pt}\lcolorbox{yellow}{s := h.next;}|
21    if (h = Head)
22     if (h = t) {
23       if (s = null)
24         return EMPTY;
25       cas(&Tail,t,s);
26     }else {
27       v := s.val;
28       |\setlength{\fboxsep}{1pt}\lcolorbox{yellow}{b:=cas(\&Head,h,s);}|
29       if(b) return v; }
30 } }
\end{lstlisting}
\end{minipage}
}
\end{narrow}
\vspace{-2ex}
\caption{The algorithm of MS lock-free queue.}\label{pic:MS}\vspace{-1ex}
\end{figure}

\begin{figure}[htpb]
\vspace{-4ex}
\includegraphics[scale=.50]{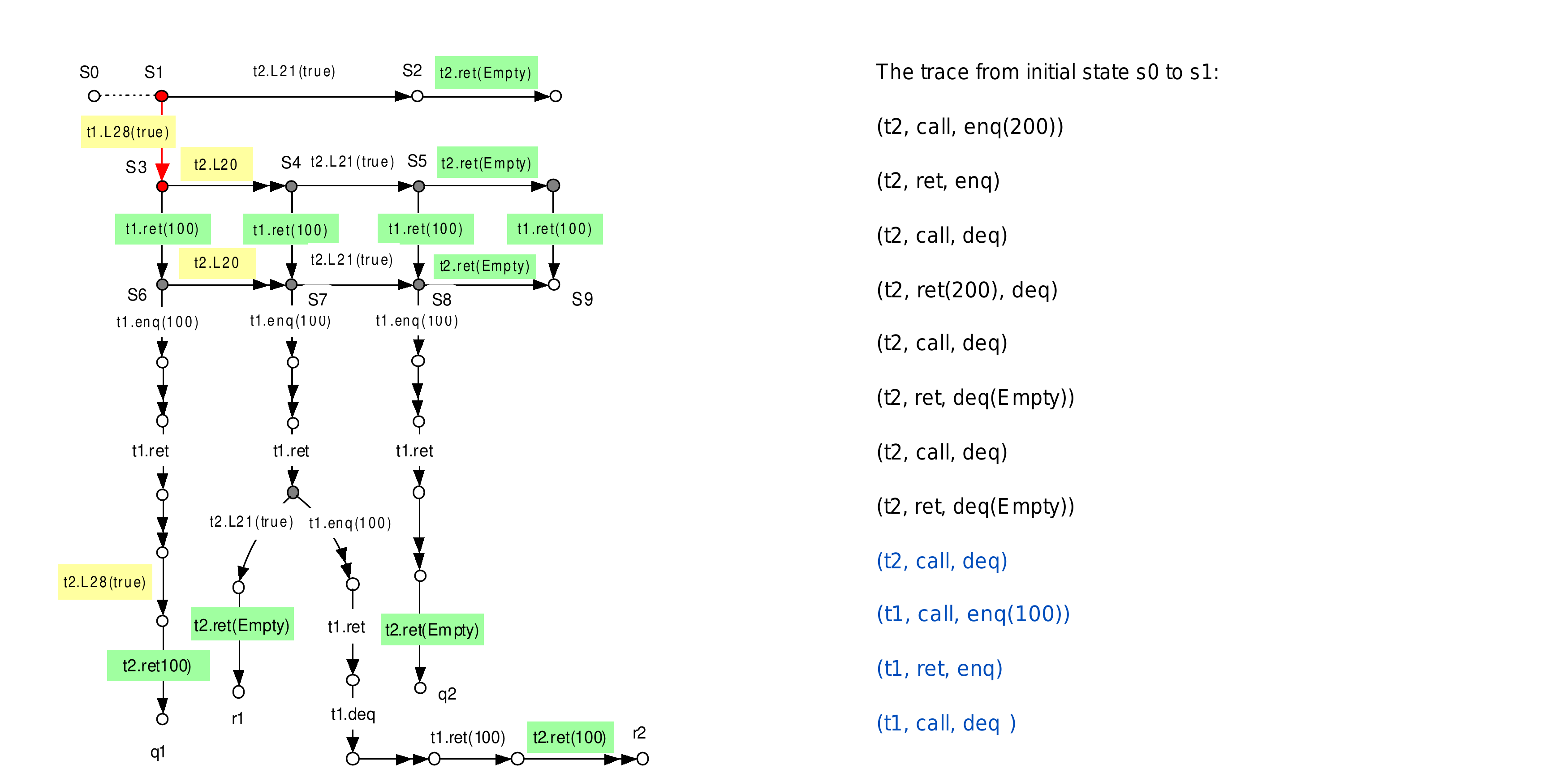}
\caption{The (part) transition system for the MS lock-free queue.}\label{pic:MS-graph}
\vspace{-.5em}
\end{figure}

Consider  a system consisting of 2 client threads, each  invoking methods \texttt{enq(v)} and \texttt{deq}  5 times.
The transition system is partly depicted in Figure~\ref{pic:MS-graph}, where
$s_0$ is the initial state, and $\twoheadrightarrow$ means
$\xLongrightarrow{}$.
The trace from $s_0$ to $s_1$ (shown in dotted line in the figure) 
is listed in text form on the right.

The transition $s_1 \xlongrightarrow{\tau(t_1.L28)} s_3$
corresponds to a successful execution of $\mathtt{cas(Head, h, s)}$
removing an element from the queue, and is a
linearization point of the call of \texttt{deq} by $t_1$.

Checking weak bisimulation with the CADP tool, it returns $s_1 \approx_w s_3$,
along with it $s_2 \not\approx_w s_4$ and  $s_2 \approx_w s_5$.
For branching bisimulation, the tool reports $s_1 \not \approx s_3$,
along with it $s_2 \not\approx s_4$ and  $s_2 \approx s_5$.

To explain the difference, consider, for instance, the transition
$s_1 \xlongrightarrow{\tau} s_2$. In weak bisimulation, it can be matched by
$s_3 \xLongrightarrow{\tau} s_4 \xlongrightarrow{\tau} s_5$,
despite that $s_2 \not\approx_w s_4$.
However, this is not allowed in branching bisimulation because
$s_2 \not\approx s_4$.